\documentstyle[sprocl]{article}
\begin{document}
\bibliographystyle{unstr}
 \title {MULTIPARTICLE DYNAMICS 1998: SUMMARY TALK} 
\author {I.M. DREMIN}     
\address { Lebedev Physical Institute, Moscow 117924, Russia    \\
E-mail: dremin@td.lpi.ac.ru \\}
\maketitle
\begin{center}
Contents
\end{center}
1. Pomeron physics\\
2. Multiplicities\\
3. Single-particle distributions\\
4. Correlations and fluctuations\\
5. Special final states and interaction mechanisms\\
6. Cosmic rays and forward detectors\\

\vspace{2mm}
Many new experimental results and theoretical ideas about multiparticle 
production processes appeared during the last year. Some of them have been
reported at this Symposium, and I try to describe the presented state of
affairs in brief, referring for more complete presentation to the original
talks and references therein. The Figures demonstrating the statements of
this review are not shown here since they can be found in the original talks.
In each section, experiment goes first followed by theory.

Compared with previous years, I notice the steady tendency of increasing 
interest in correlation studies and interactions of nuclei considered as a
strongly interacting medium, i.e. in collective effects in particle physics.
The non-perturbative aspects of multiparticle dynamics attract more attention
even though the perturbative QCD demonstrates further success in its region
of applicability.

\section{Pomeron physics}

One of the main problems we wanted to understand for a long time is whether
we are at asymptopia or have not reached it yet. In 1956, when 10 GeV Dubna
accelerator started its operation, I came as an undergraduate student to 
I. Pomeranchuk and remember him answering this question: "Yes, because
$10\gg 1$!" Actual c.m.s. energy was about 4 GeV only, compared to nowadays
1800 GeV at Tevatron and awaited 14000 GeV at LHC. On the contrary, in 1998
we are sure that the asymptopia is far away and we hardly have a chance to
come there. Our belief is based on strong increase of total cross sections 
with energy, small-$x$ behavior of structure functions and other experimental 
facts, as well as on slow (logarithmic) approach to asymptotical regime and
large higher order corrections in the perturbative QCD.

The data of ZEUS and H1 collaborations on deep inelastic scattering (Derrick,
Marage) imply the Pomeron effective intercept somewhere in the region between
1.08 and 1.2 in accordance with earlier results and expectations. The same 
conclusion follows from strong increase of the 
structure function at small $x$ and fixed $Q^2$. The scaling violation is well
fitted by QCD with next-to-leading corrections. The various distributions of
dijets in DIS favor hard $q$--hard $g$ picture.

The exclusive processes of vector meson production from $\rho $ to $\Upsilon $
have been studied at HERA in detail. The slope of the differential cross section
$b$ decreases with increase of the parton virtuality $Q^2$ as expected. Since
$b=R^2$, where $R$ is an effective distance in $q\bar{q}$-pair created by
$\gamma ^*$, it indicates the stronger color transparency effect of mutual 
screening of color charge of components of the pair at small $R$ (analogous to 
Chudakov effect in electrodynamics). The total cross section of 
$\rho $-production behaves as $\sigma (\gamma ^* \rho)\propto (Q^{2}+m_{\rho }
^{2})^{-n}$ with $n=2.24\pm 0.09$ at large $Q^2$ that is quite close to awaited
$Q^{-6}$ behavior. At smaller $Q^2$ the decrease is slower. However the overall
normalization of the cross section can be estimated theoretically (by Pomeron-
exchange models) with large uncertainty (up to the order of magnitude).
With $Q^2$ increasing, the shares of heavier mesons increase compared to
the $\rho $-share.

The results on helicity amplitude behavior show that the ratio of the 
longitudinal to transverse cross sections increases with $Q^2$ but not as fast
as theory predictions what poses a new problem to theorists. The shape analysis 
of diffractive processes reveals that the central particle density at $y=0$
in DIS is higher than in $e^{+}e^{-}$. Also, the forward-backward correlations
are stronger in DIS.

The data of HERA have been compared with the data on inelastic diffraction in
$pp$-processes at Tevatron (Goulianos) where single, double and two-Pomeron
diffractive processes as well as $W$ and heavy flavor production are analyzed.
These processes should tell us about the evolution of hadronic systems on a
large time scale, but the space-time picture is not clear yet.
The CDF data on dijets and $W$ indicate the gluon content of Pomeron of about
60-70$\%$ (with 40-30$\%$ of quarks, correspondingly). To consolidate the data
about the energy behavior of the single diffraction cross section (with a 
"knee" at $\sqrt {s}=22$ GeV), the $M^2$-dependence of its differential cross 
section and $W$-production, it has been proposed to use the renormalized flux
of Pomerons corresponding to the scaling of the gap probability. The theoretical
implications of such a procedure should be thought over. The pretentious claim
to explain at the next conference "What is Pomeron?" has been put forward, and 
we  will remember it!

The study of two-Pomeron diffraction processes at 85, 300 and 450 GeV (Kirk)
has revealed an interesting effect of the azimuthal correlation of two recoiled 
protons. It can be used to learn more about the Pomeron-Pomeron interaction.

Concerning the theoretical approach to Pomeron problem, we used to think in 
terms of the so called soft and hard Pomerons. Soft Pomerons are ascribed to 
solutions of DGLAP equations with $x$-distribution in DIS of the type
$W(x,Q^2)\propto \exp \sqrt{c\ln \frac {1}{x}\ln \frac {Q^2}{\Lambda ^2}}$.
Hard Pomeron is related to BFKL equation and gives rise to the behavior
$W(x,Q^2)\propto (x/x_0)^{-\Delta }$ where in Born approximation one estimates
$\Delta _{B}=\frac {\alpha _S}{\pi }12\ln 2 \approx 0.4-0.5$ at $Z^0$. Surely, 
the phenomenological pole-like
fits with $\alpha _{P}(0)>1$ should be considered as preasymptotical ones 
because they would violate the Froissart bound in asymptotics. In QCD,
$\alpha _S\rightarrow 0$ asymptotically, and therefore there is no direct
contradiction with the Froissart bound. Nevertheless, the high 
intercept of the hard Pomeron always looked especially suspicious, and new 
results (Lipatov) show that corrections to it are so high indeed 
($\Delta _{NL}=\Delta _{B}(1-a\Delta _{B})$ with $a=2.3$) that the position
of the intercept moves closer to 1 and becomes completely undefinite until
next corrections are calculated. Thus the clear separation between soft and
hard Pomerons has been lost in what concerns the energy behavior. Are they
mixed? Moreover, the problems with unbounded (from
below) spectrum and with oscillations (!) of total cross sections in 
$\overline {MS}$-scheme (absent in BLM-scheme ?) should be solved.

On the phenomenological side we saw impressive fits (Block) of the total, 
elastic, differential cross sections as well as ratios of real to imaginary
parts of the forward scattering amplitude in the framework of the "QCD inspired"
eikonal model. It provides $\sigma _{tot}\propto c\ln ^{2}s$ with small value of 
the factor $c$ ($\sim 10^{-3}$ of Froissart bound) and, in particular, predicts 
all above characteristics to be measured at LHC. It states that zero curvature
of the slope of $d\sigma /dt$ at Tevatron indicates that we are at preasymptotic
region yet, and its curvature should become negative at LHC. Let us see whether 
it works there, and what the chosen form of the eikonal phase means.

The soft+hard Pomeron picture (called heterotic Pomeron) has been still in use 
(Tan)
to account for minijet production at high energies. It attempts to incorporate
diffusion in the impact parameter (soft Pomeron and large color dipoles) and in 
$p_T$ or virtuality (hard Pomeron and small dipoles). The factorization seems
to be broken as it happens for multiple BFKL poles as well. It would be 
desirable to get the correspondingly modified DPM to confront such an idea 
to multiparticle production data.

To study Pomeron properties, the method of factorial moments was applied to 
final states of $P\gamma ^*$-collisions in DIS (Zhang). It is shown that the 
second factorial moment (i.e. the dispersion of multiplicities) is sensitive 
to Pomeron structure and insensitive to its flux. The relationship with above 
experimental data should be clarified.

\section{Multiplicities}

The study of multiplicity distributions (MD) is fruitful because they contain
in the integrated form all the correlations of the interacting system. The 
intensive search for scaling laws and QCD predictions for MD (they are 
infrared-safe!) give impact for the theoretical activity.

New precise experimental data on MD in $pp$-processes in the energy range from 
300 GeV to 1800 GeV of Tevatron has been reported (Walker). If compared with earlier
UA5 data at the same energy 546 GeV, it shows wider MD, is more precise at high 
multiplicities and less at low. Let us note that just these two wings of MD
correspond to unusual events with extremely developed and underdeveloped
cascades, respectively. Therefore they are especially interesting and must be
accurately measured. Both the density and the average number of
particles increase with energy. The MD maximum is located at
$n_{max}\approx 0.8\langle n\rangle $ and the shoulder is visible that provokes
to fit it by a sum of two NBD (four free parameters!) with an energy threshold
for the second NBD and about 30$\%$ of $\sigma $ at 1.8 TeV in there. It reminds 
of old conjecture of DPM (or QGSM) that the rescattering produces MD with
several shoulders. The similar shoulder has been seen in $e^{+}e^{-}$ 
processes. 

With some assumptions about the energy dependence of NBD parameters, 
one can get (Ugoccioni) predictions for MD at LHC energies from UA5 data but 
Tevatron data implies that
initial fits should be reconsidered. Also, for the fit to be more consistent,
the triple rescattering must be taken into account. 

The general question about 
foundations of NBD is usually answered in the phenomenological way. That is why
its generalizations have been used. The only one with some QCD background is
the generalized NBD (Hegyi) which stems from the Poissonian transform of MD 
obtained in some
approximation of QCD and known as the generalized gamma-distribution.
Nevertheless, even though quite effective in phenomenological fits, NBD does 
not appear in QCD which predicts different asymptotical behavior of MD moments
compared with NBD. Namely, at asymptotic energies the ratio of cumulant to
factorial moments behaves in QCD as $q^{-2}$ while in NBD with $k=2$ (this
choice is closest to QCD in $q$-asymptotics) it is represented by the
different formula $2/q(q+1)$ i.e. with twice larger ratio in asymptotics. 
Other scaling laws beside KNO have been looked for
(Ploszajczak) in the form $\langle n\rangle ^{\delta }P_{\delta }(n)=
f((n-\langle n\rangle )/\langle n\rangle ^{\delta })$ with $\delta \neq 1$
in search for a possible signature of phase transitions.

QCD predicts quite distinct oscillating preasymptotic behavior of cumulant
moments of MD (and, consequently, of their ratio to factorial moments) with
first minimum at their rank about $q_{min}\approx 5$. Experiment in $e^{+}e^{-}$
at $Z^0$ peak supports this prediction (Metzger) for the full phase space as 
was known earlier from other data. Sometimes it is interpreted as a combined 
result of hadronization and MCs showers described by NLO, not NNLO. I'd like
to stress that NNLO terms in analytical QCD appear as a byproduct of higher 
order derivatives taking into account conservation laws fully inserted
in MCs which therefore include beside NLO some contributions of higher
order approximations of QCD as well. That is why their agreement with data
is puzzling only in the sense that the hadronization does not destroy it 
even at very high ranks.

Special attention has been paid to multiplicities in gluon and quark jets
(Gary, Langefeld). Their ratio $r$ slightly exceeds 1 and increases with 
energy. According to QCD, it should approach asymptotically the ratio 
$C_{A}/C_{F}=2.25$. Experimentally at $Z^0$ peak $r=1.51\pm 0.02\pm 0.05$
for the full phase space and $r=1.82\pm 0.04\pm 0.06$ for soft particles 
($y\leq 2$). In NLO, it is equal to 2.05 and disagrees with experiment. If 
higher order corrections, the energy conservation in the triple vertices and 
the overall conservation law are taken into account
in formalism of QCD equations for generating functions, agreement can be 
restored. However, there is yet an unsettled problem with higher order
analytical calculations in the renormalization group approach which claim 
smaller corrections. 

Even more puzzling is a reported rather large difference between 
the ratio of energy slopes of average multiplicities and $r$ itself which
changes from 0.95 at lower energies to 0.7 at $Z^0$. It should be much smaller
according to theoretical estimates because it is proportional to the slope
of the running coupling constant which is small in pQCD. It demonstrates the 
running property of the QCD coupling being zero for fixed coupling.
Do we see here the non-perturbative effect at work as it has been claimed by 
DELPHI collaboration? If so, it would be very important. Before answering this
question one should understand the role of higher order QCD terms which
becomes even more essential in the slopes than in the ratio $r$. Still it is 
hard to reconcile this fact with the well known small slope of the running 
coupling constant. Besides, the asymptotical series expansion of pQCD does 
not account for the exponentially (in $\sqrt {\ln s}$) damped terms which are
important for fits starting at comparatively low energies. Theoretically, it is
easy to estimate that for the second derivatives the corresponding difference is
approximately twice larger but it seems difficult to measure it in experiment
with high enough precision. These differences are sensitive to higher 
order QCD corrections since each derivative enlarges the role of higher orders. 
Also, the energy evolution of dispersions
of MD in quark and gluon jets would be of interest. With high statistics at 
$Z^0$ and elaborated methods of jets separation it can be studied and compared 
with theory which predicts the difference in asymptotic values of 
dispersions and in their slopes. 
The progress in this field is really impressive.

Multiplicities in heavy quark jets (de Angelis) also have some special 
features. The accompanying gluon radiation of heavy quarks should be suppressed
due to their high mass compared with that of light quarks. The difference in 
multiplicities of $b\bar b$- and light quark events is predicted in QCD to stay 
constant with energy while in the "conservative" model taking into account just
energies it should decrease with energy. Earlier data up to LEP1-energies did
not allow to decide which approach is correct due to both experimental and
theoretical uncertainties. Experiments at LEP2 strongly support the QCD
conclusion. I would like to note that the angular distribution of the 
accompanying radiation in $b\bar b$-events should be of a "ring-like" or
"dead-cone" type i.e. suppressed at forward angles as determined by propagators
of $b$-quarks. It has not been studied yet in experiment.

An elaborated procedure of restoration of $\pi ^0$-MD (Krasznovszky) from the 
secondary photons in $\pi p$-processes at 40 and 250 GeV has supported the 
common belief that KNO-scaling is valid in this energy region for neutral
pions as well as for charged pions.

Finally, one should stress that the knowledge of MDs is important not only by 
itself but also for practical purposes of predicting the neutrino beams to be 
sent from CERN to Gran Sasso, for example (Bonesini).

\section{Single-particle distributions}

Single-particle distributions have been intensively investigated in many 
reactions and, especially, in high-energy nucleus-nucleus collisions during 
last years (Seyboth, Humanic, Sandor, Wang, Tserruya, Masera). The universal 
feature seen in PbPb-events at 158 GeV
for secondaries $\pi ^{\pm}, p^{\pm}, K^{\pm}, \Lambda ^{\pm}, \Xi ^{\pm }$ 
is their exponential decrease with the transverse mass $m_{T}=\sqrt {m_{0}^{2}
+p_{T}^{2}}: \; dN/dm_{T}\propto m_{T}\exp [-m_{T}/T]$, where $T$ is 
parameterized as $T=T_{0}+\beta _{t}^{2}m_{0}, \; T_{0}=140 MeV, \\
\beta _{t}
\approx 0.4, \; m_{0}$ is the mass of the particle, $p_T$ its transverse 
momentum. Let us stress that up to now the explanation of such universality
is obtained in the framework of the thermodynamical approach only. For 
multi-strange hyperons the inverse slopes $T$ deviate from the linear increase
with the particle mass.

The parameter $T$ increases for a bigger system, namely $T_{PbPb}>T_{SS}>T_{pp}$.
It is interesting to note that in $pp$-collisions there is no substantial 
dependence of $T$ on $m_0$. Sometimes this nuclear effect of increase of the
inverse slope with particle masses is ascribed to the transverse radial flow 
of particles or to the final-state rescattering. As seen from these universal 
distributions, the integrated yields are also larger for a bigger system i.e.
$Y_{PbPb}>Y_{SS}>Y_{pp}$. For particles of the same species but with different
electric charges the integrated yields are different. For example, the ratio of
$K^+$ to $K^-$ integrated yields in PbPb at 158 GeV is about 1.84 (larger than 
in SS or $pp$) and increases for lower energies. It is of interest in 
connection  with a possible signature of the quark-gluon plasma in 
event-by-event analysis.

The proton rapidity distributions from central PbPb-collisions show the 
phenomenon of baryon stopping i.e. of comparatively large fraction of protons 
in the central rapidity region. Fritiof model predicts smaller baryon stopping,
while RQMD overestimates it. Multiple collisions could qualitatively explain
baryon stopping. At the same time, the pion rapidity distribution in the whole
rapidity range ($0<y<6$) is insensitive to baryon stopping. It scales from SS
to PbPb and peaks at midrapidity.

The strangeness enhancement has been observed in nucleus-nucleus collisions 
and shown to increase with the strangeness content of a particle. It is often
considered as a signature of possible phase transition to quark-gluon plasma.

Very intriguing observations have been reported (Stassinaki, Perepelitsa,
Tserruya, Masera) about the production of photons
and dileptons in $(\pi ,K, p)p$, SS, PbPb collisions at energy of hundreds GeV.
In the region of $\rho ,\omega $-resonances the strong excess (over all 
traditional calculations) of photons with 
low transverse momenta $p_{T}^{(\gamma )}<40$ MeV has been seen as well as the 
low $p_T$ excess for $e^{+}e^{-}$ pairs with masses in the range $0.25<
m_{e^{+}e^{-}}<0.68$ GeV. The excess factor 
ranges from 3 to 7 compared with QED and decay estimates in different reports.
It depends on the rapidity range studied, is stronger in the central region 
and  increases with particle density at midrapidity. Attempts to explain it 
include hypothesis about $\pi \pi $-annihilation, shift of masses and widths
of $\rho ,\omega $ with the temperature as well as more exotic ones like 
appearence of special domains in the QCD vacuum. However, no reliable
conclusion has been reached. Something happens at large distances, and we have 
no explanation. Are wee partons in charge of this effect? If yes, it would be
a signature of a non-perturbative mechanism. Can Bose-Einstein effects
reveal its origin? Still, there is no signal 
consistent with the large excess of low-mass electron-positron
pairs in SAu photon measurements and in pBe at 450 GeV by HELIOS collaboration.

In the region of high-mass dilepton pairs, one observes the well known 
anomalous $J/\psi $
suppression in PbPb so that the $E_T$ integrated cross section is reduced by 
the factor $0.74\pm 0.06$ with respect to the absorption model prediction. 
The ratio of dimuon decays of $J/\psi $ to Drell-Yan pairs shows strong decline 
for PbPb from usual exponential fit valid for lighter colliding nuclei and 
hadrons. There is some enhancement in the mass region between $\rho ,\omega $ 
and $J/\psi $ which could be ascribed to enlarged 
charm production. An additional $\psi ^{'}$ suppression is observed both in
SU and PbPb data.

The detailed impressive review of efforts of 4 collaborations preparing for 
future experiments at RHIC has been presented (Jacak). Our general hope is 
that the experimental results on interaction of objects with high complexity
beside revealing their geometrical structure will give us insight into the
properties of the QCD vacuum, in the equations of state of the hadronic matter, 
in specifics of field theories at the finite temperature. They will help us
understand the properties of the heavy neutron stars and of the matter at
early stages of the Universe.

The distributions in DIS and $e^{+}e^{-}$ processes have been compared and 
discussed in connection with QCD predictions (Milstead, Chliapnikov, B\"{o}hrer,
Sarkisyan). The status of the hump-backed plateau in $\xi =\ln 1/x$-variable
is confirmed with the position of the peak and its width well fitted by MCs
both in DIS and $e^{+}e^{-}$. The reported approximate scaling of 
$\xi _{max}/\langle n\rangle ^{1/2}$ is somewhat of surprise since in QCD
$\xi _{max}\propto \ln s +O(\ln ^{1/2}s)$, while $\langle n\rangle ^{1/2}
\propto \exp [c\ln ^{1/2}s]$ so that it can approximately hold just in some
energy interval only. It must be sensitive to the angular ordering 
(contrary to Chliapnikov's statements)
because otherwise $\langle n\rangle $ drastically differs for ordered and 
non-ordered cascades (the factor $c$ is different). The scaling violations
in $ep$ are well described by pQCD but the event shape spectra need power 
corrections.
The particle content in $ep$  and 3-jet $e^{+}e^{-}$ events 
with different species studied poses some problems because MCs are sometimes
unable to describe it.

The high mass and high $p_T (E_T)$ physics at Tevatron with further plans of
the accelerator reconstruction have been described (Giokaris). The relative 
yields, cross sections and masses of heavy mass states with limits on Higgs
masses (about $<250$ GeV) are presented. The controversial issue of the excess
at very high $E_T$ claimed by CDF has been discussed in connection with the 
quark  compositeness problem.

\section{Correlations and fluctuations}

This topic is naturally the most widely discussed in multiparticle dynamics.
It can be separated in three parts dealing, correspondingly, with correlations
of identical particles (HBT-effect), general correlations and fluctuations 
studied mostly by moments behavior, and special collective effects and cascade
modifications. There were many talks devoted to it. The 
integrated form of correlations appears already in the overall multiplicity 
distributions as has been separately discussed. The recent conference 
"Correlations and Fluctuations" in Hungary was specially devoted to this subject,
and more detailed presentations can be found in its proceedings.

The main aim of studies of correlations between identical particles (HBT-effect)
which arise due to (anti)symmetrization of the wave function is to understand
the space-time parameters of the interaction (Smirnova, de Jong, Humanic, Wang). 
Such a symmetrization results in
excess over 1 of the normalized correlation function for two identical bosons 
in the region of small relative momenta with a width inverse proportional to 
the radius of the interaction region. For identical fermions (usually, two
protons) the antisymmetrization gives rise to the negative correlation function
at very small momenta, changing sign and possessing a positive maximum which
indicates their interaction range. Both curves measured in nucleus-nucleus
collisions for $\pi \pi $ and $pp$ agree with our intuitive ideas about 
nuclei sizes. However somehow they show that while the production of particles
in nucleon collisions is effective at distances $r\leq 1.5$ fm, it 
becomes crucial at much larger distances $r>5$ fm in nucleus-nucleus collisions.
It would imply some retardation effect within the future light cone for the
collision process. It depends on the transverse masses and 
atomic number decreasing for larger $m_T$ as $m_{T}^{-1/2}$ and increasing for 
heavier nuclei. The attempts to explain $m_T$-dependence by rescattering or
flow are questionable because the similar effect has been observed in 
$e^{+}e^{-}$. Is the dynamics of this long-distance effect related somehow
to large time scales essential for diffraction processes?
It is interesting to note that the peak height in $pp$-correlations (equal
about $1.14\pm 0.04$) does not depend on energy in the wide energy range from
100 MeV to 200 GeV where physics of the process must change drastically.

In AA, there is no dramatic difference between the longitudinal and transverse sizes
for various species and at high transverse momenta (with somewhat larger values 
of $R_{long}$ for low $k_T$ pions only). Therefore it is hard to explain the
abovementioned $J/\psi $ suppression by the drastic change of the space-time
picture. Such a difference has been noticed 
in $e^{+}e^{-}$ and hadron-hadron collisions. In $e^{+}e^{-}$, the shape of the 
region is elongated so that $R_{long}\approx 2R_{out}\approx 2R_{side}$.
With overall fit of the form $C_{2}=N(1+\lambda _{2}\exp [-R^{2}Q^{2}])$ it has
been found that $R\approx 0.5$ fm, $\lambda _2$ increases and $R$ decreases 
with $m_T$ increasing (compare to AA, where $\lambda _2$ does not increase 
while $R$ decreases as well). With increase of the multiplicity, the radius
$R$ increases also. One should remind, however, that the variable $Q^2$ is 
not very well suited for such fits, and, in practice, the points at small $Q^2$
are usually omitted when fits are done.

The three-particle correlations in $e^{+}e^{-}, \pi p, pp$ differ
from AA (Bialas, Smirnova, Sarkisyan).
The 3-particle cumulant moments revealing "genuine" correlations differ from 
zero in $e^{+}e^{-}$ (with $R_{3}\approx 0.6$ fm) and within experimental errors 
are consistent with zero in AA. Let us note, however, that the amplitude of 
oscillations of $H_q$-moments increases for targets with more complex structure.
It shows that higher rank cumulants may become larger just in AA.

Interesting conjectures have been made about the role of HBT-correlations in
processes of $W$-production in $e^{+}e^{-}$-collisions. Various theoretical 
schemes predicted that the difference between the charged multiplicities in 
processes with both produced $W$'s decaying in hadrons and the doubled 
multiplicity of those processes when only one of the two created $W$'s 
decays hadronically $\Delta (n_{ch})=n_{\pi }(WW)-2n_{\pi }(W)$ should be
non-zero and range between $-3$ and 2.1 (Fialkowski) due to  HBT-correlations.
The latest experimental data (de Jong) at $\sqrt s$=183 GeV (LEP2) show 
$\Delta _{exp}=0.2\pm 0.5$ and no difference in $x$-distribution of the two 
processes. However, the preliminary DELPHI data at 172 GeV gives $\Delta =
3.9\pm 1.6$. The influence of HBT-correlations on the mass shift of $W$
was earlier estimated sometimes as exceeding 50 or even 100 MeV but recent
(model dependent) results show that it ranges between 0 and $50\pm 20$ MeV.
Therefore one is tempted to conclude that no clear signal of influence 
of HBT-correlations
on $W$ masses has been seen until now. The importance of the color 
reconnection in $W$-mass shifts can be clarified by studying  the "inclusive"
3-jet events (Gary).

The problem of implementation of HBT-correlations in MC-schemes is non-trivial 
due to the quantum-mechanics origin of this effect. It has been addressed by
three groups. Two of them (Todorova-Nova, Lund) deal with matrix elements. 
The third group (Fialkowski) deals with the density matrix and iterative 
rescaling procedure with Gaussian BE-functions in JETSET. First results claim
that $R_{long}\neq R_T, \;  K_{3}\neq 0 \; $ and $\Delta M_{W}\sim 10\pm 10$ 
MeV that agrees with discussion above. Some cross-checks should be still done
(e.g. on the positivity of the Wigner function in the last approach).

The study of general correlations and fluctuations in smaller phase space 
regions is usually related (Bialas) to notions of intermittency and fractality.
Started for hadronic processes, it has been extended to all initial particles
and nuclei. Actually it shows  the evolution of MD for ever smaller phase 
space  regions in 1, 2 or 3 dimensions. According to QCD predictions in DLA 
(sometimes, with some NLO corrections taken into account as well) the 
factorial moments should first increase linearly on the double-logarithmic scale
for smaller intervals with subsequent "knee" and curvature at ever smaller
intervals. Qualitatively, these trends are seen in experimental data (Mandl).
The approximation used for analytic calculations is too crude to be 
quantitatively true. In the computer solution of equations for generating 
functions with conservation laws properly taken into account, the agreement
is very good both for moments in small bins and in the whole phase space region
(oscillations of $H_q$-moments discussed above) as was shown (Lupia) at recent
"Correlations and Fluctuations'98" conference. I would comment that the overall 
energy conservation leads to power corrections, and they can contribute several
per cents at comparatively low energies. However no power-like dynamical terms 
have been taken into account. Probably, more important fact is that in computer
calculations no Taylor expansion is used, and therefore all high order terms
are included. MC models are also able to fit 
the experimental data reasonably well. Thus I do not see any important 
disagreement here but would prefer to rely on the qualitative features and
on limitations of the method than insist on quantitative description now. 
The only but crucial problem is that, with one-to-one
correspondence between partons of pQCD and hadrons, it looks as if the
boundary between short-distance and large-distance phenomena has been 
shifted so far that no space for large distances is left. Is it so?
How does it correspond to new conclusions from multiplicities of quark and 
gluon jets?

Further multidimensional extension of this method uses the 
difference between the scales in different directions of the phase space and,
therefore, the self-affine transformations (Liu). The values of respective
Hurst exponents show the degree of the anysotropy in different directions.
It has been also proposed (Blazek) to exploit other characteristics such as
frequency moments and dispersion moments. These proposals must be applied to
a wide range of experimental data to prove their fruitfulness.

In DIS, the correlations between the current and target regions in the Breit 
frame have been studied (Chekanov) by measuring the function $\rho =\langle 
n_{c}n_{t}\rangle -\langle n_c\rangle \langle n_t\rangle $ which is predicted 
to be negative in QCD. It is sensitive to the gluonic structure function and to
gluon density, however, the hadronization influences it. The forward-backward 
correlations have been also measured and the anticorrelation was observed with
$\rho _{FB}=\langle n_{F}n_{B}\rangle -\langle n_{F}\rangle \langle n_{B}\rangle
\approx -2.5$ in the range $10<Q^{2}<10^{3}$ GeV$^2$. For identical colliding 
particles, it must be negative in a sample of
events with a fixed multiplicity. Thus its sign is not a surprise, and it would 
be desirable to know how much of its value is due to dynamics and mixture of
different multiplicities, respectively.  

Many important theoretical questions concerning correlations have been raised
at the Symposium: Are there collective effects where particles and nuclei act
as a medium? Are there specific correlations modifying the parton cascade? How
can we observe phase transitions if any? etc. For example, if the nucleus still
"remembers" during the initial stage of collision that it is quite a solid
object, then "sling effects" could be noticed. Not discussed here was 
the widely disputed problem of the suppression of gluon radiation in a
hadronic medium (analogous to the so called Landau-Pomeranchuk-Migdal effect
in QED).

The influence of the hadronic medium on final correlations can be observed
in predicted peculiar correlations of pions with the opposite sign charges
slowly moving in opposite directions with equal (in modulus) momenta (Andreev).
The origin of this effect is the fact that fields in a medium are not free and
undergo Bogoliubov transformation to become free at the end. Therefore the
particle and antiparticle operators are mixed and the "shadow" of 
HBT-correlations appears in that form.

The predicted a year ago and only recently published (Andersson) effect of
screwiness at the end of the parton cascade has been already confronted to
experiment (de Angelis). The correlation length in $p_T$ and helicity 
conservation in a 3-gluon vertex define a special angular distance so that the 
dipoles can not be very short, and the gluonic classical field acquires a helix 
structure as if screwed on the phase-space cylinder. 
It would produce a special pattern of events in the $(\theta, \phi )$-
plane with some periodicity or, in other words, display a peak at the "resonant"
value of frequency $\omega $ in an expression $\vert \sum _{j}
\exp (i(\omega y_{j}-\phi _{j}))\vert ^{2}$ with sum over rapidities $y_j$
and azimuthal angles $\phi _j$ of all particles in an event sample. 
No such peak (periodicity) has been found both in DELPHI data and in the 
traditional JETSET73-based simulation. It does not exclude the chance 
that the period is either shorter or longer than those which
could be resolved. Other possibilities are that the resonance decays smear it
or the multiplicity is still too
low to get a clear effect in event-by-event analysis and the effect is washed
out in a sample of events. Further studies are needed.

Such a long discussed collective effect specific for hadron-nucleus collisions
as the cumulative effect has been described in the framework of the "QCD 
inspired" model (Braun). It consists in recoil of impinging protons in the 
region kinematically inaccessible for proton-proton collisions. It has been shown
that the nuclear structure function is steeply decreasing in this region as
$\exp [-16x]$, and particle production with $\vert x\vert >1$ is damped as 
well: $\exp [-6x]$. The number of particles increases as $n\propto A^{1+\frac
{x}{3}}$, and the ratio of different species is $n_{p}:n_{\pi }:n_{K}=100:10:1$.

When discussing nucleus-nucleus collisions, I omitted several comments in 
different talks on a very important problem of existence of the phase 
transition from the hadronic phase to the quark-gluon plasma. Here I'd like to
mention the thermodynamical approach (Diakonos) which deals with this problem by
considering the chiral field $(\sigma ,\vec{\pi })$ in 3 dimensions. It predicts
clusters of pions with intermittent behavior in the momentum space which give
 rise to low-multiplicity critical events with minijet-like structure in
$(y, \phi )$-plane. The second order phase transition is advocated. The whole
approach is similar to considering the O(4) Heisenberg magnet in 3 dimensions.

Some collective effects deserved, in my opinion, more attention than just the
private discussions. Among them are the energy flows, the peculiarities in
pseudorapidity distributions of dense and narrow groups of particles, the
abovementioned LPM and sling effects.

The methods of experimental study of correlations evolve nowadays to studying
the patterns in high-energy and high-multiplicity individual events. The 
progress in this direction
would be noticeable, in particular, if one is able to measure the impact 
parameter in each event. It has been claimed (Treleani) that the large number
of minijets in heavy ion collisions at LHC energies and a very narrow 
distribution
at $p_{T}^{min}\rightarrow 0$ will allow to fix the impact parameter in a
subsample. The geometrical picture with multiparton distributions has been used.
In my opinion, the difference in virtualities of colliding partons can widen
the distribution. Therefore, the stability of conclusions concerning the impact 
parameter with respect to varying virtualities of partons creating minijets
should be proved and various $x$-regions considered. 

If this method is just a 
proposal for future studies, the method of the wavelet analysis finds already 
real applications. Earlier, discrete wavelets were used for analysis of some 
theoretical cascade models. I proposed to use the continuous wavelets for
pattern recognition in individual high-multiplicity events, and EMU-15 data
were used. At this Symposium, the discrete wavelets have been used for that
same purpose (Petridis) with analysis of NA49 data on PbPb interactions at
158 GeV as well as of some theoretical and MC models. It helps reveal the
substructures on the event-by-event basis at various scales and is more 
powerful than the traditional Fourier analysis. One hopes to provide the
classification scheme when using this method and to find out events with
special patterns (jets, minijets, DCC, ring-like events, azimuthal flows,
screwiness, sling effects etc). It is especially important in search for 
limitations of theory and of simulation tools.

\section{Special final states and interaction mechanisms}

The spectroscopy of the final state particles was among the main topics a 
decade ago. Nowadays the PDG tables are filled so densely that main interests 
shifted to mechanisms of particle production. There were several experimental
(Shephard, Kirk, Walker, Sandor, Tserruya, Masera, Chliapnikov, Bohrer, 
Giokaris) and theoretical (Calucci, Draggiotis, Wong, Musulmanbekov, Treleani,
Diakonos, Xie Qu-bing) talks where these problems were touched (sometimes in
combination with topics discussed above). Without describing them in more 
detail, I will briefly mention these problems just to show their variety and 
refer to original talks for more details. Spectroscopy interests move now to
exotic mesons, glueball searches, Higgs, sparticles or to the partial wave 
analysis e.g. looking for exotic quantum numbers $J^{PC}=1^{-+}$ signals
in such final states as $\eta \pi ^{0},\; \eta \pi ^{-},\; \rho \pi ^{-},\; 
\eta ^{'} \pi ^{-}$ etc
(Shephard). In nucleus-nucleus interactions, the most hot discussions are about
the existence of the quark-gluon plasma. In this direction, the search for 
strangeness enhancement, strangelets, $J/\psi $ suppression, dileptons, DCC, 
color rearrangement, jets propagation in nuclei, rescattering, direct photons
and many other effects is going on. Production rates of $d$ and $\bar {d}$ are
essential also for comparison with their ratio in our Galaxy (Walker). In 
$e^{+}e^{-}$-collisions, there were reported at this Symposium results on
inclusive production at $Z^0$, on high masses and high-$p_T$ events  with 
special emphasis on jet properties, on some exotic channels with extremely low
cross sections like $e^{+}e^{-}\rightarrow Z^{0}Z^{0}\rightarrow b\bar {b}c\bar
{c}$ etc. Some new models of the hadron structure with relationship between 
constituent and current quarks, of hadroproduction with an ambitious program 
of computation of multiparticle amplitudes in the framework of the effective 
action approach and the generating function technique, and of a Hamiltonian 
model for such processes have been reported as well but they are at the very
initial stage and can attract the attention only after their internal problems 
are solved and the successful application to experiment done what has not been
attempted until now.

\section{Cosmic rays and forward detectors}

The review talks on cosmic ray physics (Schmitz, Stanev, Jones, Lindner,
Fonseca, Lorenz) were of interest especially in connection with new physics 
results and RHIC and LHC coming to operation. Surely, the recent discovery
at SuperKamiokande of neutrino oscillations and neutrino masses attracted
much attention.

Concerning the multiparticle dynamics, it has been stressed again that the 
energy region covered by cosmic ray studies ranges from $10^{11}$ to $10^{20}$
eV in the lab. system, and accelerators actively enter this field with Tevatron
at $2\cdot 10^{15}$eV and LHC at $10^{17}$eV. The cosmic ray fluxes at these 
energies are very low already. However, there is the substantial difference in 
the regions of the phase space studied in CR and at accelerators. The 
fragmentation region with main energy flux and relatively small number of 
secondary particles is investigated in CR installations while LHC detectors
will see mostly the pionization region with many comparatively slow particles
(90$\%$ of the energy flux will be outside the ATLAS detector!). Leading 
particle properties carry parton information and could provide better 
discriminants than structure functions. Probably, BFKL effects are enhanced
there.

New facilities for future CR experiments have been described. They include huge
installations covering several thousands of squared kilometers!

Many facts from CR observations which have not been seen at accelerators
attract the attention of particle physicists:

1. A {\it knee} of the CR spectrum at energies about $10^{15}-10^{16}$ eV can
be of interest if it is due to big changes in the nature of generic particle 
interactions
at that energy as it is speculated sometimes. It is more trivial if it happens 
due to the change of the CR flux composition.

2. {\it Centauros} are exotic events with many charged hadrons created and no
(or small number) gammas from neutral pion decays. Respectively, antiCentauros
contain many gamma-rays and no (or small number) charged particles.
Recent idea of the 
disoriented chiral condensate with the anomalous distribution of $\pi ^0$
compared with $\pi ^{\pm}$ of the type $P(f)\propto f^{-1/2},\; f=n_{0}/n_{tot}$
favors Centauros. However, more simple explanations with isospin conservation
in coherent or squeezed states would do the same job. Recent results of CR
Brazilian group again show that 20-30$\%$ of DCC is needed. However, CDF 
results show that $\sigma _{Cent}\leq 10\mu b$ (within $1.3<\vert \eta \vert
\leq 4.1$), and T864 (MiniMax) experiment claims that there is no Centauro
signal at few per-cent level albeit in the central pseudorapidity region as 
well.

3. {\it MiniCentauros} - no signature in accelerator experiments yet.

4. {\it Chirons} are miniclusters with very small relative $p_T$. Are these
fluctuations in core distributions of the dynamical origin?

5. {\it Heavy flavors}. JACEE collaboration has studied in detail 15 events 
from the low-multiplicity part of MD with less than 50 secondaries at the 
primary energy higher than 1 TeV per nucleon. 
64 secondary vertices consistent with $b$ or $c$ decays have been found.
Can one explain this peculiarity of the low-multiplicity events?
Is it due to the energy increase of the heavy flavor production cross section?

6. {\it Long-flying cascades} (Tien-Shan experiment) at $E>10 ^{14}$ eV 
penetrate deeply in the installation. They could be explained if the cross 
section of charm hadroproduction increases with energy and becomes about 
1.5 mb at $10^{14}$ eV. I reported about it 12 years ago at the similar
Symposium. Also, strangelets have been considered as a possible 
explanation.

7. {\it Aligned events} (Pamir experiment) are characterized by the non-uniform
distribution of the high-energy secondaries and cores in the emulsion chamber
so that they tend to be placed along straight lines. There were 6 multicore 
events observed. The attempts to explain them by a string decay result in 
estimates which are twice below their production rate.

8. {\it Inelasticity} i.e. the share of energy going to secondary non-leading
particles $K=\langle 1-\frac {E'}{E_0}\rangle $ decreases with energy 
increasing as favored by CR data but disfavored by most models.

9. {\it Violation of Feynman scaling} in the fragmentation region is claimed
according to analysis of CR data. The common problem for CR calculations is the 
uncertainties with recalculation of $pp$ to $pA$ and $AA$ processes and with 
mass composition of primary CR radiation.

Thus we have many CR facts in the fragmentation region of multiparticle 
production which apart Pomeron diffractive processes urgently ask for doing 
the forward physics at accelerators. Some forward detectors are built at RHIC.
Unfortunately, most operating and almost all planned detectors are not aimed
at these problems. Nevertheless there are some proposals of detectors which 
cover the forward region (only the last one in this list was described here
but I mention also some others and the physicists dealing with
these proposals):

1. FELIX detector (Bjorken et al) has not been approved yet.

2. CDF proposal for forward detector (Goulianos) also has not been approved
until now.

3. Emulsion chambers with any target (Kotelnikov) can be easily installed and 
operated as fixed target (lower energies!) experiments in an extracted beam
(Tevatron, HERA, LHC?). Usual problem with low statistics can be soon overcome
with the help of automatic tables for emulsion processing installed.

4. CASTOR detector (as a forward detector of ALICE at LHC) has been described 
(Bartke) at this Symposium. Operating at LHC with the pseudorapidity coverage 
$5.6\leq \eta \leq 7.2$ it would check the CR results on Centauros, long-flying
component and strangelets, in particular. 

However, this detector has been only
partly approved and it will start operating at LHC not earlier than in 2005
while some above proposals can be put forward quite soon, in principle, albeit 
at lower energies. We hope that physics community will get financial resources 
for these proposals to be at work in the nearest future and we will learn 
many more interesting facts about multiparticle dynamics at next meetings.

\section*{Acknowledgments}
On behalf of all the participants, I want to thank Nikos Antoniou and all our 
hosts for a very stimulating meeting and for such a memorable visit, in general.
I am grateful to organizers of the Symposium for inviting me to give this talk
and for financial support. This work was also supported by the Russian Fund for
Basic Research (grant 96-02-16347) and by INTAS.

\end{document}